\documentstyle[aps]{revtex}
\begin{document}
\tightenlines
\title{Quantally fed steady-state domain distributions\\
in Stochastic Inflation}
\author{Mauricio Bellini\thanks{E-mail: mbellini@mdp.edu.ar}, Pablo D.
Sisterna\thanks{E-mail:sisterna@argenet.com.ar} and Roberto R.
Deza\thanks{E-mail: deza@mdp.edu.ar}}
\address{Departamento de F\'{\i}sica, Facultad de Ciencias Exactas y
Naturales,\\ Universidad Nacional de Mar del Plata,\\ De\'an Funes
3350, 7600 Mar del Plata, Argentina.}
\maketitle
\begin{abstract}
Within the framework of stochastic inflationary cosmology we derive
steady-state distributions $P_c(V)$ of domains in comoving
coordinates, under the assumption of slow-rolling and for two specific
choices of the coarse-grained inflaton potential $V(\Phi)$.  We model
the process as a Starobinsky-like equation in $V$-space plus a
time-independent source term $P_w(V)$ which carries
(phenomenologically) quantum-mechanical information drawn from either
of two known solutions of the Wheeler-De Witt equation:
Hartle-Hawking's and Vilenkin's wave functions.  The presence of the
source term leads to the existence of nontrivial steady-state
distributions $P_c^w(V)$.  The relative efficiencies of both
mechanisms at different scales are compared for the proposed
potentials.
\end{abstract}
Since the differential microwave radiometer (DMR) mounted on the COBE
satellite first detected temperature anisotropies in the cosmic
background radiation, we have the possibility to directly probe the
initial density perturbation.  The fact that the resulting energy
density fluctuations (${\frac{\delta\rho}{\rho}}\approx.5\times
10^{-5}$) fit the scaling spectrum predicted by the inflation model,
suggests that they had indeed their origin in the quantum fluctuations
of the ``inflaton'' scalar field during the inflationary era.
Although this problem is in principle of a quantum-mechanical nature,
the fact that under certain conditions (which are made precise in
Ref.\cite{BCMS}) the inflaton field can be considered as classical
largely simplifies the approach, by allowing a {\em Langevin\/}-like
stochastic treatment.  {\em Stochastic Inflation\/} branched off and
grew up from within the context of {\em chaotic inflation\/}
\cite{GL}.  It was Starobinsky \cite{Starobinsky} the first one to
derive, from a Langevin-like equation for the dynamics of the
coarse-grained inflaton field $\Phi$ and in the absence of
back-reaction from the metric, a Fokker-Planck equation for the
transition probabilities $P_c(\Phi,t|\Phi',t')$ in comoving
coordinates.  The latter provide us with statistical information about
the relative number of spatial domains that evolve in a time interval
$t-t'$ from having a typical value $\Phi'$ (assumed constant
throughout the domain) towards a new configuration with a typical
value $\Phi$.  This approach is purely phenomenological and keeps very
little information about its quantum origin, since the coarse-graining
procedure erases quantum correlations.  Hence, the issue of
integrating quantum-mechanical information to the inflationary
evolution is still a pending assignment.

In a recent paper \cite{BSD} we have offered a phenomenologically
drawn example that takes into account---if not properly
back-reaction---at least the possibility that Hubble's constant $H$
depend on $\Phi$.  Our approach disregarded instead the boundary
conditions at Planck's and exit-of-inflation $\Phi$ regimes and, in
particular, domain injection from the purely quantum-mechanical phase.
As we know, the Wheeler-De Witt (WDW) equation is a possible way to
describe the fully quantum-mechanical behaviour of the Universe.
However, this equation has infinitely many solutions: hence, there
have been several attempts to solve it by starting from different
initial conditions, each providing a different mechanism whereby
domains with a typical size $a=H^{-1}$ can enter the inflationary 
stage.  Two very popular ones among them are {\em quantum creation
from ``nothing''\/} \cite{Vi} and the {\em no boundary proposal\/}
\cite{HH}.

It is our aim in this work to investigate the inflationary stage in
the light of the information provided by solutions to the WDW
equation.  Our (still phenomenological) approach will be to enlarge
the scope of Starobinsky's equation by adding a time-independent
source term.  We shall take this (extrinsic) domain-injection
probability per unit time to be proportional to the squared modulus of
the aforementioned solutions.  Moreover, we shall assume that during
the inflationary phase the slow-rolling condition is satisfied.  Under
these assumptions, we obtain phenomenologically meaningful stationary
domain distributions in $V$-space for two physically appealing forms
of the inflaton potential $V(\Phi)$.

For the role that quantum cosmology is intended to play in this work
(i.e.\ to act as an effective source of homogeneous horizon-sized
domains) it suffices to consider the {\em minisuperspace\/} version of
the WDW equation \cite{linde}:
\begin{equation}\label{WdW}
    \left[-(3\pi M^2_p)^{-1}\frac{\partial^2}{\partial a^2}
    +\frac{3\pi M^2_p}{4}a^2+(2\pi a)^{-2}
    \frac{\partial^2}{\partial\Phi^2}-2\pi^2a^4V(\Phi)\right]
    \Psi(a,\Phi)=0
\end{equation}
This is the cornerstone for a quantum-mechanical description of the
pre-inflationary era, and is derived under the assumption that the
metric {\em is\/} of the Friedmann-Robertson-Walker (FRW) type
\begin{equation}\label{FRW}
    ds^2=\sigma^2\left[N^2(t)dt^2-a^2(t)d\Omega_3^2\right]
\end{equation}
and that {\em the only\/} quantum variables in this stage are the FRW
scale factor $a$ and the inflaton field $\Phi$.  In eq.(\ref{FRW})
$N(t)$ is the lapse function, $d\Omega_3^2$ the metric on a unit
three-sphere and $\sigma^2=2G/3\pi$ a normalizing factor chosen for
convenience.  Equation (\ref{WdW}) arises as a {\em constraint\/} when
varying the Hamiltonian with respect to $N$.

The asymptotic solution of eq.(\ref{WdW}) with the boundary condition
$\Psi(a\to\infty)\rightarrow 0$ was obtained by Vilenkin \cite{Vi}, in
analogy with the decay of a metastable state.  Taking
$a=H^{-1}(\Phi)$, its squared modulus
\begin{equation}\label{Vi}
    P_v(V)=\exp{(-3M_p^4/8V)}
\end{equation}
describes the {\em nucleation\/} of universes (domains in our
phenomenological model) with a given value of the inflaton potential
$V$ in the inflationary phase {\em through quantum tunneling}, and the
corresponding process is known as ``quantum creation from nothing''.
On the other hand, the celebrated Hartle-Hawking (HH) solution
\cite{HH} leads to
\begin{equation}\label{HH}
    P_h(V)=\exp{(3M_p^4/8V)}.
\end{equation}

We find that eq.(\ref{Vi}) and eq.(\ref{HH}) differ in the sign of the
exponential.  In the following---and unless otherwise specified---we
shall denote the domain-injection probability arising from any
solution to the WDW equation as $P_w(V)$.  Obviously, given a specific
functional form $V(\Phi)$ one immediately obtains $P_w(\Phi)$.

Now we want to incorporate this information into the equation for
$P_c(V,t)$ (the probability distribution of domains in comoving
coordinates during the inflationary stage).  After choosing units such
that $M_p=1$, changing variables in the Fokker-Planck equation for
$P_c(\Phi,t)$ and recalling that ${\frac{\partial}{\partial\Phi}}=V'
{\frac{\partial}{\partial V}}$ (the prime denotes $\frac{d}{d\Phi}$),
one obtains
\[
\frac{\partial}{\partial t}P_c^w(V,t)=\sqrt{\frac{2}{3\pi}}V'
\frac{\partial}{\partial V}\left[V^{\frac{3}{2}(1-\beta)}V'
\frac{\partial}{\partial V}\left(V^{\frac{1}{2}\beta}P_c^w\right)
+\frac{3V'}{8V^{1/2}}P_c^w\right]
+C^w\,P_w(V),
\]
where the time-independent domain distribution $P_w(V)$ acts as a
source term that accounts for domain injection into the inflationary
phase, and $C^w$ is a (unknown) constant.  $\beta=1/2$ for the
Stratonovich prescription and $\beta=0$ for It\^{o}'s one.

In this work we shall restrict ourselves to the {\em stationary\/}
equation which, by defining
\begin{equation}\label{14}
    G_w(V)=V^{\frac{3}{2}(1-\beta)}V'\frac{d}{dV}\left(V^{\frac{3}{2}
    \beta}P_c^w\right)+\frac{3V'}{8V^{1/2}}P_c^w,
\end{equation}
can be rewritten as
\begin{equation}\label{ec}
    \frac{d}{dV}G^{sol}_w(V)=-C_1^w\,\frac{P_w(V)}{V'}
\end{equation}
with $C_1^w={6\sqrt{6\pi}}C^w$, and $G^{sol}_w(V)$ is the function of
$V$ solution of the last equation.  Once we solve eq.(\ref{ec}) we can
obtain $P_c^w(V)$ from eq.(\ref{14}), setting $\beta=1$ as in
Ref.\cite{LLM}:
\begin{equation}\label{ei}
    P_c^w(V)=K_o\,V^{-3/2}\exp{\left(\frac{3}{8V}\right)}\left[B+\int\,
    dV\,\exp{\left(-\frac{3}{8V}\right)}\frac{G_w^{sol}(V)}{V'}\right].
\end{equation}
Here $B$ is an integration constant and $K_o=V_e^{-1}-1$ (with
$V_e=V(\Phi_e)$) comes from normalizing at the {\em exit-of-inflation\/}
value $\Phi_e$ of the inflaton field.

In the remaining of this letter---and with the aim to illustrate the
procedure---we derive the steady-state domain distributions $P_c^v(V)$
and $P_c^h(V)$ arising from both domain-injection mechanisms, for two
specific choices of $V(\Phi)$.
\begin{enumerate}
    \item Let us first consider the case of a quadratic potential $V
    (\Phi)=\frac{m^2}{2}\Phi^2$, so that $V'=\left(2m^2V\right)^{1/2}$:
    \begin{enumerate}
	\item If the domains are produced by quantum tunneling,
	$P_w(V)$ is Vilenkin's distribution eq.(\ref{Vi}).  For
	simplicity we choose $C_1^w=-1$.  This gives
	\[
	G^{sol}_v(V)=\frac{1}{m}\left[\frac{1}{\sqrt{2V}}
	\exp{\left(-\frac{3}{8V}\right)}+\frac{1}{2}\sqrt{3\pi}
	{\bf{\rm Erf}}\left(\sqrt{\frac{3}{8V}}\right)\right] ,
	\]
	where ${\rm Erf}$ is the error function.  The general solution
	of eq.(\ref{ei}) is
	\begin{equation}
	    P_c^v(V)=K_o\,\exp{\left(\frac{3}{8V}\right)}
	    \left[B_v^{(1)}-f_1(V)\right],
	\end{equation}
	where $B_v^{(1)}$ is a constant of integration and
	\[
	f_1(V)=-\frac{1}{m}\int dV\,\exp{\left(-\frac{3}{8V}\right)}
	\left[\frac{1}{\sqrt{2V}}\exp{\left(-\frac{3}{8V}\right)}
	+\frac{1}{2}\sqrt{3\pi}{\bf{\rm Erf}}
	\left(\sqrt{\frac{3}{8V}}\right)\right].
	\]
	From the requirement that $P_c^v(V)$ be {\em positive in the
	inflationary regime\/} it follows that $B_v^{(1)}\ge
	|\min[f_1(V)]|$.  We need another condition in order to fully
	specify the integration constants.  Considering our ignorance,
	we will require both solutions to {\em agree at the Planck
	scale} i.e.\ that $P_c^v(V_p)=P_v(V_p)$, where
	$V_p=V(\Phi_p)$.  Then
	\[
	B_v^{(1)}=\frac{e^{-81}}{V_e-1}+f_1(V_p).
	\]
	\item In the case of a HH source term \cite{HH} and choosing
	$C_1^h=1$, it results
	\[
	G^{sol}_h(V)=-\frac{1}{m}\left[\frac{1}{\sqrt{2V}}
	\exp{\left(-\frac{3}{8V}\right)}+\frac{1}{2}\sqrt{3\pi}
	{\bf{\rm Erf}}\left(\sqrt{\frac{3}{8V}}\right)\right].
	\]
	The steady-state domain distribution in comoving coordinates
	is then
	\begin{equation}
	    P_c^h(V)=K_o\exp{\left(\frac{3}{8V}\right)}
	    \left[B_h^{(1)}-g_1(V)\right],
	\end{equation}
	where
	\[
	g_1(V)=\frac{1}{m}\int dV\,\exp{\left(-\frac{3}{8V}\right)}
	\left[\frac{1}{\sqrt{2V}}\exp{\left(-\frac{3}{8V}\right)}
	-\frac{1}{2}\sqrt{3\pi}{\bf{\rm Erf}}\left(\sqrt{\frac{3}{8V}}
	\right)\right]
	\]
	and $B_h^{(1)}$ is an integration constant required now to be
	$B_h^{(1)}\ge|\min[g_1(V)]|$.  Imposing as before
	$P_c^v(V_p)=P_v(V_p)$,
	\[
	B_h^{(1)}=\frac{1}{V_e-1}+g_1(V_p).
	\]
    \end{enumerate}
    The functions $f_1(V)$ and $g_1(V)$---describing the domain flow
    from the quantum sector towards the inflationary regime---must be
    evaluated numerically in this example.
    \item The following is an exactly solvable example:
    \begin{enumerate}
	\item Let us consider in equation (\ref{ec}) the choice
	$P_v(V)/V'=V^{-2}$, and choose also $C_1^w=-1$.  The resulting
	scalar potential is
	\[
	V(\Phi)=\frac{3}{8\ln\left[\frac{3}{8}(\Phi_p-\Phi)\right]}.
	\]
	The condition $V(\Phi)\geq 0$ imposes the restriction
	$\Phi_p-\Phi\geq 8/3$.  Since, on the other hand
	$G^{sol}_v(V)=-V^{-1}$, it is possible to find $P_c^v(V)$ from
	equation (\ref{ei}):
	\begin{eqnarray}
	    P_c^v(V)&=&K_oV^{-3/2}\exp{\left(\frac{3}{8V}\right)}
	    \nonumber\\
	    &\times&\left[B_v^{(2)}-\exp{\left(-\frac{3}{8V}\right)}
	    V^{1/2}\left(\frac{2}{3}V-\frac{1}{2}\right)
	    +\frac{\sqrt{6\pi}}{8}\,{\bf{\rm Erf}}
	    \left(\sqrt{\frac{3}{8V}}\right) \right] .
	\end{eqnarray}
	Here $B_v^{(2)}$ is again a constant ensuring that the
	distribution $P_c^v(V)$ be positive in the whole inflationary
	domain.  By choosing $P_c^v(V_p)=P_v(V_p)$ and taking $V_p=1$,
	it results to be
	\[
	B_v^{(2)}=\frac{V_e}{\left(1-V_e\right)}+\frac{1}{40}
	\left[23\,e^{3/8}-\sqrt{6\pi}{\bf{\rm Erf}}
	\left(\sqrt{\frac{3}{8}}\right)\right].
	\]

	\item When the source term is HH-like, choosing $P_h(V)/V'=
	V^{-2}\exp{\left(\frac{3}{4V}\right)}$---so that $G^{sol}_h(V)
	=\frac{4}{3}\exp{\left(\frac{4}{3V}\right)}$---the solution
	for $P_c^h$ (with $C_1^h=1$) results:
	\begin{eqnarray}
	    P_c^h(V)&=&K_oV^{-3/2}\exp{\left(\frac{3}{8V}\right)}
	    \nonumber\\
	    &\times&\left[B_h^{(2)}+\frac{1}{5}
	    \exp{\left(\frac{3}{8V}\right)}V^{1/2}\left(\frac{8}{3}V^2
	    +\frac{2}{3}V+\frac{1}{2}\right)
	    -\frac{\sqrt{6\pi}}{40}\,{\bf{\rm Erf}}
	    \left(\sqrt{\frac{3}{8V}}\right)\right].
	\end{eqnarray}
	By choosing $P_c^v(V_p)=P_v(V_p)$,
	\[
	B_h^{(2)}=\frac{V_e}{\left(1-V_e\right)}e^{-3/4}
	-\frac{1}{6}e^{-3/8}+\frac{\sqrt{6\pi}}{8}{\bf{\rm Erf}}
	\left(\sqrt{\frac{3}{8}}\right),
	\]
    \end{enumerate}
\end{enumerate}

Figure 1 compares the ratio $Z(V)=P_c^v(V)/P_c^h(V)$ as a function of
$V$ for the two cases, between the values $V_e\approx 0$ and $V_p=1$.
Although both curves are rather similar in form when looked in their
own scale, $Z(V)$ for the logarithmic potential looks very flat when
compared with the quadratic one.  Moreover---except for $V<0.1$---the
quadratic potential predicts larger values of $Z(V)$.  In fact,
whereas for the quadratic potential and $V>0.1$ it is $Z(V)>1$ (with
a maximum larger than 3 at $V\approx 0.3$)---thus implying a higher
efficiency of the {\em nucleation\/} mechanism as compared to the HH
one---for the logarithmic one it is always $Z(V)<1$ (thus implying a
higher efficiency of the HH mechanism) and this relative efficiency
passes through a (less dramatic) minimum at $V\approx 0.2$.

Summing up, in this work we have studied the effect of the
domain-injection mechanism on the (stationary) domain distribution at
the end of the inflationary era, for two phenomenologically appealing
proposals of inflaton potential forms: {\em quadratic\/} and {\em
logarithmic\/}, respectively.  The domain-injection mechanisms are
provided in our approach by known solutions to the WDW equation.
Although the ratio $Z(V)$ remains of order one and looks similar for
both potential forms, it is concluded that the relative efficiency of
each mechanism depends on the form of the inflaton potential.

This letter is a first step towards studying the possible appearance
of stability in Stochastic Inflation due to quantum sources of
inflationary domains, deserving the matter further research and
debate.  We are aware of the lack of naturalness of our approach, that
combines such different descriptions as quantum cosmology and
(gravitationally classical) stochastic inflation.  The transition from
quantum to classical behaviour in quantum cosmology, and particularly
the conditions under which the wave function of the Universe becomes
semiclassical, are far from being understood.  Thus, considering the
possibility that the exit of the quantum phase is not a well defined
event, a stochastic interplay between both classical and quantum
descriptions seems plausible.

\begin{figure}[tbp]
\caption{The ratio $Z(V)=P_c^v(V)/P_c^h(V)$ between the steady-state
distributions arising from Vilenkin's and Hartle-Hawking's source
terms, for the case of a quadratic potential (upper curve) and for a
logarithmic one (lower curve).}
\label{fig1}
\end{figure}

\begin{references}
\bibitem{BCMS} M. Bellini, H. Casini, R. Montemayor and P. Sisterna,
Phys.\ Rev.\ D {\bf 54}, 7172 (1996).
\bibitem{GL} A. S. Goncharov and A. D. Linde, Sov.\ J. Part.\ Nucl.\
{\bf 17}, 369 (1986).
\bibitem{Starobinsky} A. A. Starobinsky, Phys.\ Lett.\ B {\bf 117},
175 (1982).
\bibitem{BSD} M. Bellini, P. Sisterna and R. Deza, Nuovo Cim.\ B {\bf
113}, 471 (1998).
\bibitem{Vi} A. Vilenkin, Phys.\ Rev.\ D {\bf 50}, 2581 (1994).
\bibitem{HH} J. B. Hartle and S. W. Hawking, Phys.\ Rev.\ D {\bf 28},
2960 (1983).
\bibitem{linde} see A. D. Linde, {\em Particle Physics and
Inflationary Cosmology\/} (Harwood, 1990), ch.10.
\bibitem{LLM} see A. D. Linde, A. D. Linde and Mezhlumian, Phys.\
Rev.\ D {\bf 49}, 1783 (1994).
\end{references}
\end{document}